# Vacuum encapsulated synthesis of 11.5 K NbC superconductor

Rajveer Jha and V. P. S.Awana*

*Quantum Phenomena and Application Division, National Physical Laboratory (CSIR)*
*Dr. K. S. Krishnan Road, New Delhi-110012, India*

## Abstract

Bulk polycrystalline NbC samples are synthesized through solid state reaction route in an evacuated sealed quartz tube. Studied NbC samples are crystallized in NaCl-type cubic structure with space group *Fm-3m*. To control cell parameters and minute un-reacted phases, different samples are synthesized with various heat treatments. Finally phase pure NbC is achieved. The grain size of the as systemized material being seen from SEM (scanning electron microscopy) is non-uniform of around 3-10μm size. Crystal structure and lattice parameters of samples have been calculated by Rietveld analysis of room temperature X-ray powder diffraction data. The lattice parameter increases with synthesis temperature and scales with superconducting transition temperature ($T_c$). Both AC and DC magnetization exhibited highest $T_c$ at around 11.5 K for an NbC sample with lattice parameter a = 4.471 Å. The lower critical field ($H_{c1}$) and irreversibility field ($H_{irr}$) measured at 3 K are around 250 Oe and 4.5 kOe respectively. The upper critical field ($H_{c2}$) being determined from in-field AC susceptibility measurements is 7.8 kOe and 11.7 kOe with 50% and 90% diamagnetism criteria, respectively.

*Corresponding Author: Dr. V.P.S. Awana, (E-mail: awana@mail.nplindia.ernet.in)
Senior Scientist, NPL, New Delhi-12, India
Web page- www.freewebs.com/vpsawana/
**Tel.** +91-11-45609357, **Fax.** +91-11-45609310

# I. Introduction

The race for the discovery of new superconductors got a boost in early 1986 with invention of cuprate high $T_c$ superconductors (HTSc) [1]. Further, when the momentum was a bit slowed down, in year 2002 the discovery of $MgB_2$ [2] steamed the race once again. More recently, the scientists are once again puzzled with the invention of pnictide superconductors [3]. Interestingly, before cuprates [1], the superconductivity for over 75 years was mostly confined to the electron-phonon mediated interactions [4]. On materials science front itself, most of exotic superconductors have appeared only after the discovery of cuprates [1]. In fact after discovery of superconductivity in Cuprates [1] and Fe based pnictides [3], one wonders as if, still more surprises are yet to come from periodic table.

In this short communication, we focus on an inter-metallic superconducting compound i.e., NbC, on which there are only few scant reports [5-9] in literature. Our interest on this compound is two fold, first though it was discovered long back in 1964 itself [5], the same is yet least studied superconductor. Second, the solid state chemistry of different allotropes of Carbon in any crystalline compound had been of immense interest in last couple of decades not only for chemists but to physicists as well. Couple of examples in this regards are Fullerenes [10], Carbon Nano Tubes (CNT) [11,] and Graphene [12]. We synthesized the intermetallic Nb based Carbide NbC compound via a simple vacuum encapsulation technique. The vacuum encapsulation technique based high temperature synthesis of compounds recently gained popularity after discovery of Fe based pnictide superconductors. The superconducting critical temperatures of superconducting intermetallic compounds can be empirically related to parameters such as crystal structure, composition, free electron concentration, and lattice parameter [6-8]. It is difficult to synthesis NbC because transition metal react at high temperature. Niobium Carbide is a refractory ceramic with a high melting point (3600°C), excellent mechanical properties, and low electrical resistance [5]. Additionally, NaCl-type transition metal carbides MC (M =V, Ti, Nb, Ta, Hf and Zr) are of fundamental importance in solid-state science and technology [5-9]. Some researchers have reported the unit-cell parameter and crystal structure of these Carbides [6, 8]. Seemingly, the cell parameters and Carbon content in NbC play deciding role in deterring the superconducting transition temperature ($T_c$) [6]. Actually the lattice parameter depends upon the C content of the NbC system as it possesses wide region of homogeneity (from $NbC_{1.00}$ to $NbC_{0.72}$) [13, 14]. The disordered state of the non-stoichiometric $NbC_x$ has B1 structure. Neutron diffraction studies showed that with the C content of 0.81 to 0.88 ordered phases of $M_6C_5$-type are formed in NbC system [13]. Furthermore, the atomic displacement parameters of the Carbides have rarely been reported. Cubic NbC is a superconductor below 12 K [5-9]. In other words, innovative approaches are needed to synthesize NbC with controlled microstructures/defects so that enhanced superconducting properties, reduced electrical resistivity in the normal state at low temperatures, and improved mechanical strengths can be accomplished. From a technological point of view, this material is interesting, because of its possible application to induce superconductivity in Carbon nanotube junctions [9]. Here we report the synthesis temperature optimization and host of superconducting properties for the stoichiometric NbC. We optimized the synthesis temperature for NbC and achieved a single phase compound without any detectable impurity within the resolution limit of X-ray powder diffraction. The resultant compound is superconducting at 11.5 K.

## II. Experimental

All the studied polycrystalline NbC samples were prepared through single step solid-state reaction route via vacuum encapsulation technique. High purity (~99.8%) Nb, and nano-C in their stoichiometric amount were weighed, mixed and ground thoroughly using mortar and pestle. The mixed powders were palletized and vacuum-sealed ($10^{-4}$ Torr) in a quartz tube. These sealed quartz ampoules were placed in box furnace and heat treated at (a) 1150$^o$C for 24 hours, (b) 1200$^o$C for 24 hours and (c) at 1250$^o$C for 24 hours each. Finally furnace was allowed to cool down to room temperature naturally. The crystal structure was analyzed by the powder X-ray diffraction patterns at room temperature using Rigaku X-ray diffractometer with Cu-$K_\alpha$ radiation. The scanning electron microscopy (SEM) image of the sample (synthesized at 1250°C) was taken on a ZEISS-EVO MA-10 scanning electron microscope. The magnetic (DC and AC magnetization) measurements are carried out on the physical property measurement system (PPMS-14T) from Quantum Design-USA.

## III. Results and discussions

Figure 1 shows the room temperature XRD patterns of various NbC samples being synthesized at different temperatures. The XRD patterns of 1150°C, 1200°C and 1250°C processed samples are shown one over the other in Fig. 1. The XRD data are fitted using Rietveld refinement (FullProf Version). It is clear that samples are phase pure and crystallized in cubic structure with space group *Fm-3m*. Coordinate positions for the atoms are Nb: 1a (0, 0, 0), and C: 1b (0.5, 0.5, 0.5). We observed that with an increase in sintering temperature from 1150 °C to 1250°C, the lattice parameter is increased monotonically from 4.468 Å to 4.471 Å, see Table 1. For further higher temperature synthesized samples, the increase in lattice parameter was not observed; also the phase purity of the compound also gets compromised. Hence we show the XRD data only for up to 1250$^{o}$C synthesized sample.

Figure 2 represents the DC magnetization (M-T) plots for samples synthesized at various temperatures. Both the Zero-field-cooled (ZFC) and field-cooled (FC) data are shown. The superconducting transition onset $T_c$ is found to be 9.09 K, 11.0 K, and 11.5 K respectively for 1150°C, 1200°C, and 1250°C synthesized samples. The 1200°C sample exhibited a two step kind of transition in ZFC magnetization. It seems samples possess both the 1150°C, and 1250°C phase within. Inset of Fig. 2 depicts the $T_c$ vs lattice parameter variation plot for variously synthesized NbC samples. It can be seen that with increasing lattice parameter the $T_c$ is increasing monotonically. We synthesized NbC for higher temperatures as well but no further increase in lattice parameter is observed, also phase purity of compound gets compromised. The increase in $T_c$ of NbC with its c-parameter is in agreement with an earlier report [6]. It is reported that with increase of synthesis temperature the Carbon content in the crystal structure of NbC increases [15, 16], similar to $MgCNi_3$ [17, 18]. Since Carbon is a light element, so it is difficult to extract the exact Carbon content by XRD refinement. The exact content of Carbon can be determined through Neutron diffraction pattern, as done for $MgCNi_3$ [18]. The dependence of $T_c$ with Carbon content is reported earlier [15, 16]. On the basis of these reports the real Carbon content in the sample having $T_c$ = 11.5 K (synthesized at 1250$^{o}$C) can be estimated to be ~ 0.99. The C content in the samples synthesized at lower temperatures (1150$^{o}$C and 1200$^{o}$C) decreases and thus the $T_c$. Decrease in both the electron-phonon interaction parameter and the un-renormalized density of electron states causes this rapid decrease in $T_c$ with decreasing C concentration [8]. The structural changes occurring in nonstoichiometric ($NbC_x$) as a result of the ordering of Carbon atoms and vacancies lead to a reconstruction of their electron and phonon spectra. Earlier it is observed that the ordering of Carbon atoms and vacancies results in decreased magnetic susceptibility with an increase in the lattice parameter of the B1 basic structure [13]. In any case we found that best/optimum $T_c$ of around 11.5 K is obtained for 1250 $^oC$ samples. Neither lower (<1250$^oC$) nor higher (> 1250$^oC$) temperature synthesis could help in further improving upon the observed $T_c$ of 11.5 K. Hence now on words we focus on superconductivity characterization optimum $T_c$ (11.5 K) sample, which is synthesized at 1250$^oC$.

Figure 3(a) shows the DC magnetization of the phase pure NbC sample (Synthesized at 1250°C) in both field cooled (FC) and zero field cooled (ZFC) situations at 10 Oe field. The sample shows bulk superconductivity with an onset temperature ($T_c$ onset) of 11.5 K. Figure 3(b) and 3(c) showing the real and imaginary part of AC magnetization for the same. It is clear that both DC and AC susceptibility magnetization exhibit superconducting transition below 11.5K. The imaginary part of AC susceptibility exhibit clear single peak at around 11 K. In case of superconductors, the real part of susceptibility depicts diamagnetic shielding of the sample, while imaginary part indicates the hysteric losses due to vortex motion. Maximum loss takes place when the magnetic flux lines just penetrate the centre of the sample and is indicated by the peak in the magnetization curve at a particular temperature called as peak temperature. The single sharp peak in imaginary AC susceptibility indicates the better coupling of the grains in the studied NbC. As far as DC susceptibility and real part of AC susceptibility are concerned, we define the superconducting transition onset temperature $T_c$ at a temperature where finite change of diamagnetic moment takes place. The $T_c$ onset for the presently studied bulk NbC is at 11.5 K in both DC and AC magnetization. There seems to be large irreversibility in the DC magnetization plot, indicating towards good pinning. This compound seemingly possesses no grain boundary contributions. This explains its single intra-grain transition with the possibility of large irreversibility field.

Figure 4 represents the AC magnetization of the same (1250 $^0C$ synthesized) sample, confirming the bulk superconductivity at around 11.5 K. Further, AC magnetic susceptibility measurements have been done at 333 Hz and varying amplitude of 3-15 Oe. With change in amplitude from 3-15 Oe, the imaginary part peak height is increased along with increased diamagnetism in real part of AC susceptibility. This is usual for a superconductor. The interesting part is that the imaginary part peak position temperature (11.5 K) is not changed at all with increase in AC amplitude. This shows that the superconducting grains are well coupled and hence nearly no grain boundary contribution in the superconductivity of NbC. To check the grain connectivity and morphology of the sample we have made the Scanning Electron Microscopy (SEM) study. Figure 5 represent the SEM image of the sample synthesized at 1250$^{o}$C. Well connected grains can be seen and/or grain connectivity is such that the grain boundary contribution is almost zero. It is clear from AC susceptibility (imaginary part) results that either the grain boundaries are non-existent in NbC or transparent to the current. This is similar to that as observed in case of $MgCNi_3$ [17].

Figure 6, shows that real part of AC susceptibility being measured applied DC bias field of up to 30 KOe, and its inset depicts the isothermal magnetization (M-H) plot at 3 K in high field range of up to 20 kOe in four quadrants for optimized (1250$^{o}$C) NbC. The lower critical field ($H_{c1}$) and irreversibility field ($H_{irr}$) are around 250 Oe and 4.5 kOe, respectively, at 3 K, see inset Fig. 6. The upper critical field ($H_{c2}$) being determined from in-field AC susceptibility measurements, is 7.78 kOe and 11.73 kOe with 50% and 90% diamagnetism criteria, respectively. NbC belongs to the dirty-limit superconductor [19]. The value of $H_{c2}$ being measured from the in field *AC* susceptibility measurements is close to the paramagnetic limit $H_P = 1.84T_c$. Thus Zeeman pair breaking mechanism will not be effective in this case. It is clear that irreversibility field ($H_{irr}$) and upper critical field ($H_{c2}$) both are of the same order of magnitude.

## IV. Conclusion

Vacuum encapsulate NbC samples are synthesized and found to crystallize in cubic structure with space group *Fm-3m*. Increase in lattice parameter is observed with increase in synthesis temperature which leads to increase in C content in the system and thus the enhanced $T_c$. An optimized $T_c$ of 11.5 K is observed for the 1250°C synthesized NbC sample. AC magnetization with field and SEM image indicated good grain connectivity for studied NbC sample. The upper critical field ($H_{c2}$) being determined from in-field AC susceptibility measurements, is above 11.73 kOe with 90% diamagnetism criteria. It is clear that we succeeded in synthesis of single phase 11.5 K NbC superconductor by quartz vacuum encapsulation in a single step.

## Acknowledgements

The authors are grateful for the encouragement and support from Prof. R. C. Budhani (Director NPL) for his motivation and discussions. Rajveer Jha would like to thank the CSIR for providing the SRF scholarship to pursue his Ph.D. This work is also financially supported by Department of Science and Technology (DST-SERC) New Delhi, India.

**Table 1: Variation of Lattice parameters and transition temperature of NbC samples with synthesis temperature.**

| Synthesis Temperature(°C) | Lattice parameter (Å) | $T_c$ (K) | $\chi^2$ |
|---|---|---|---|
| 1150 | 4.468 (2) | 9.09 | 3.27 |
| 1200 | 4.469 (3) | 11.0 | 3.28 |
| 1250 | 4.470 (2) | 11.5 | 3.94 |

# Figure Captions

**Figure 1** Rietveld fitted *XRD* pattern of NbC samples with space group *Fm-3m* synthesized at different temperatures (1150°C, 1200°C, and 1250°C).

**Figure 2** DC magnetization in *ZFC* (zero-field-cooled) and *FC* (field-cooled) situation for all samples at 10 Oe.

**Figure 3** (a) DC magnetic susceptibility M(T) in ZFC and FC situations at 10 Oe of NbC, (b) real part of AC susceptibility, measure at 333 Hz and 10 Oe of NbC and (c) imaginary part of AC susceptibility, measured at 333 Hz and 10 Oe of NbC, synthesized at (1250°C).

**Figure 4** AC magnetic susceptibility in both real (*M*') and imaginary (*M*) situations at fixed frequency of 333 Hz and varying amplitudes of 3-15 Oe for NbC Synthesized at (1250°C).

**Figure 5** SEM image of NbC synthesized at 1250°C, shows well connected grains.

**Figure 6** Isothermal magnetization *(MH)* for real part of AC susceptibility (*M'*) with applied field up to 30 kOe at 3 K for NbC Synthesized at (1250°C) the upper critical field ($H_{c2}$) is marked. Inset shows expanded MH plots at 3 K in the low field range of up to 10 kOe in four quadrants for NbC.

**Figure 1**

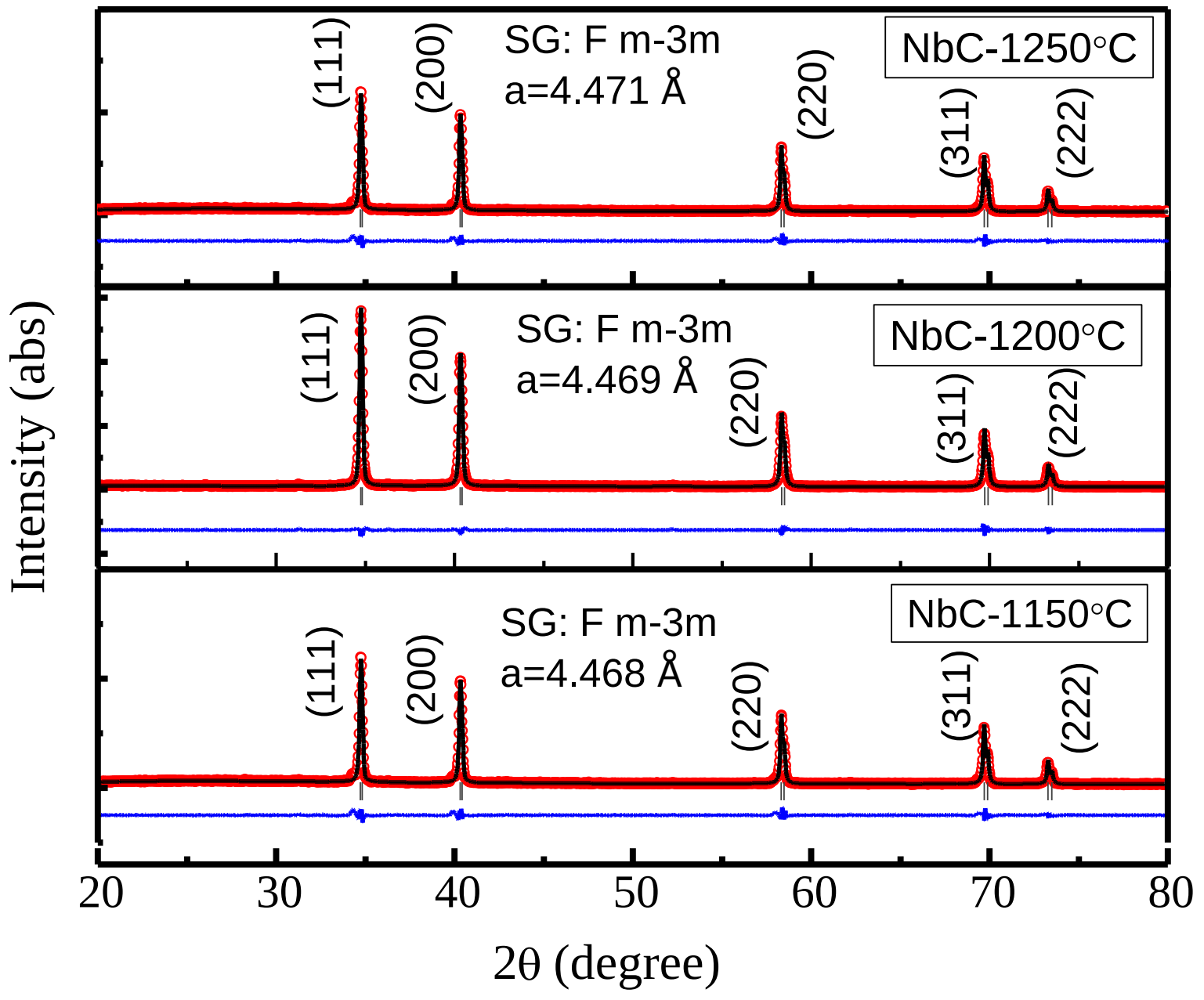

**Figure 2**

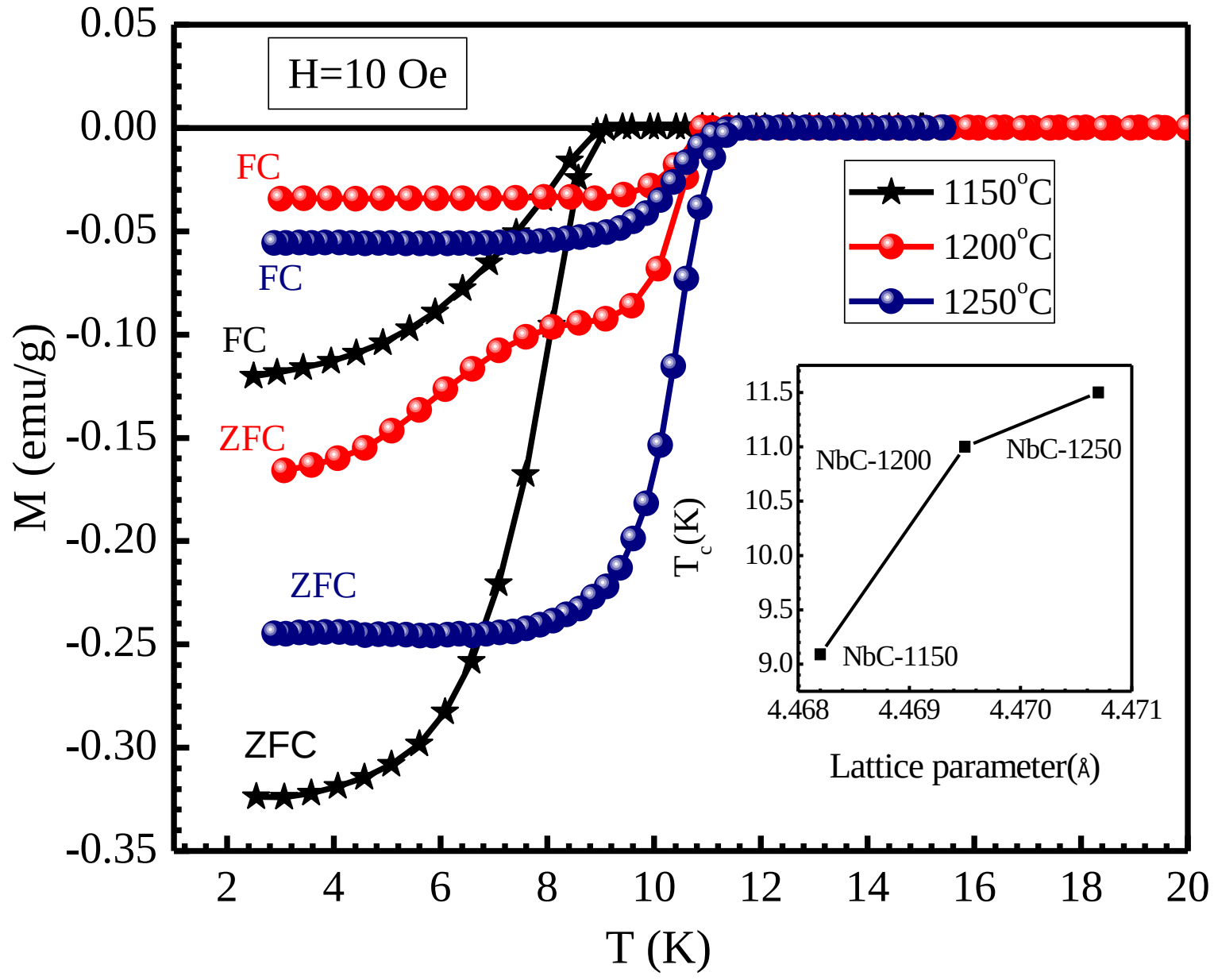

**Figure 3**

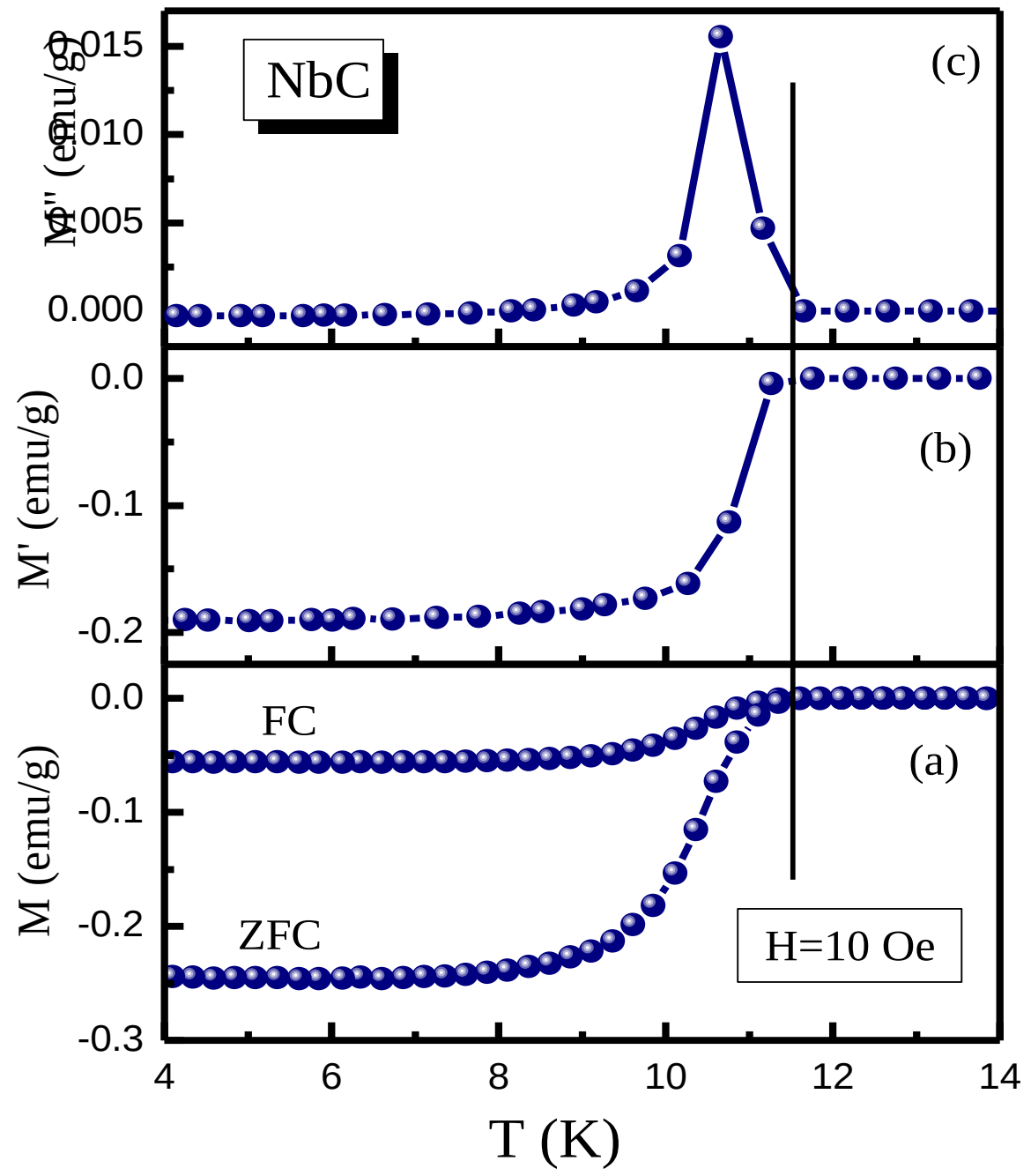

**Figure 4**

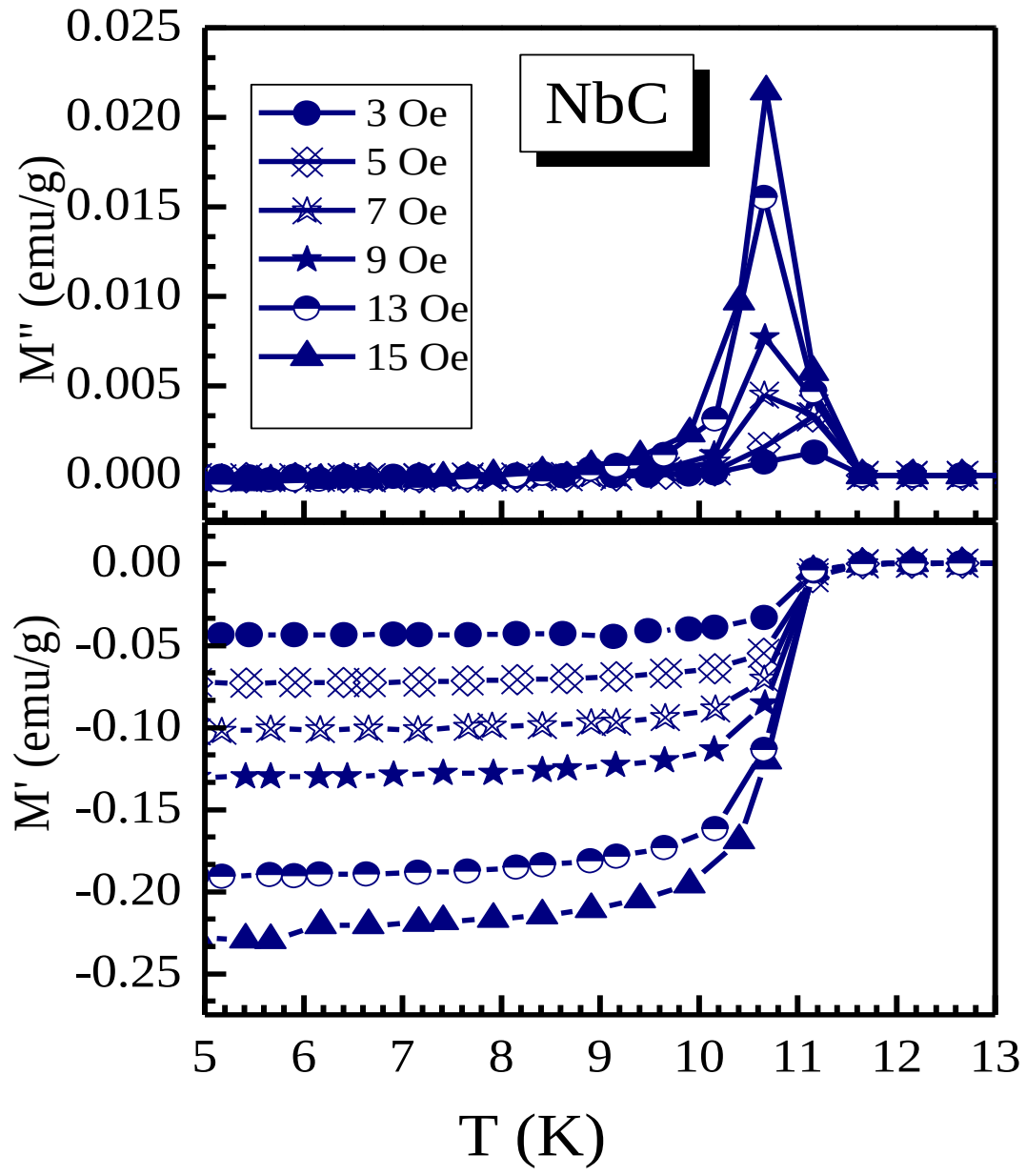

**Figure 5**

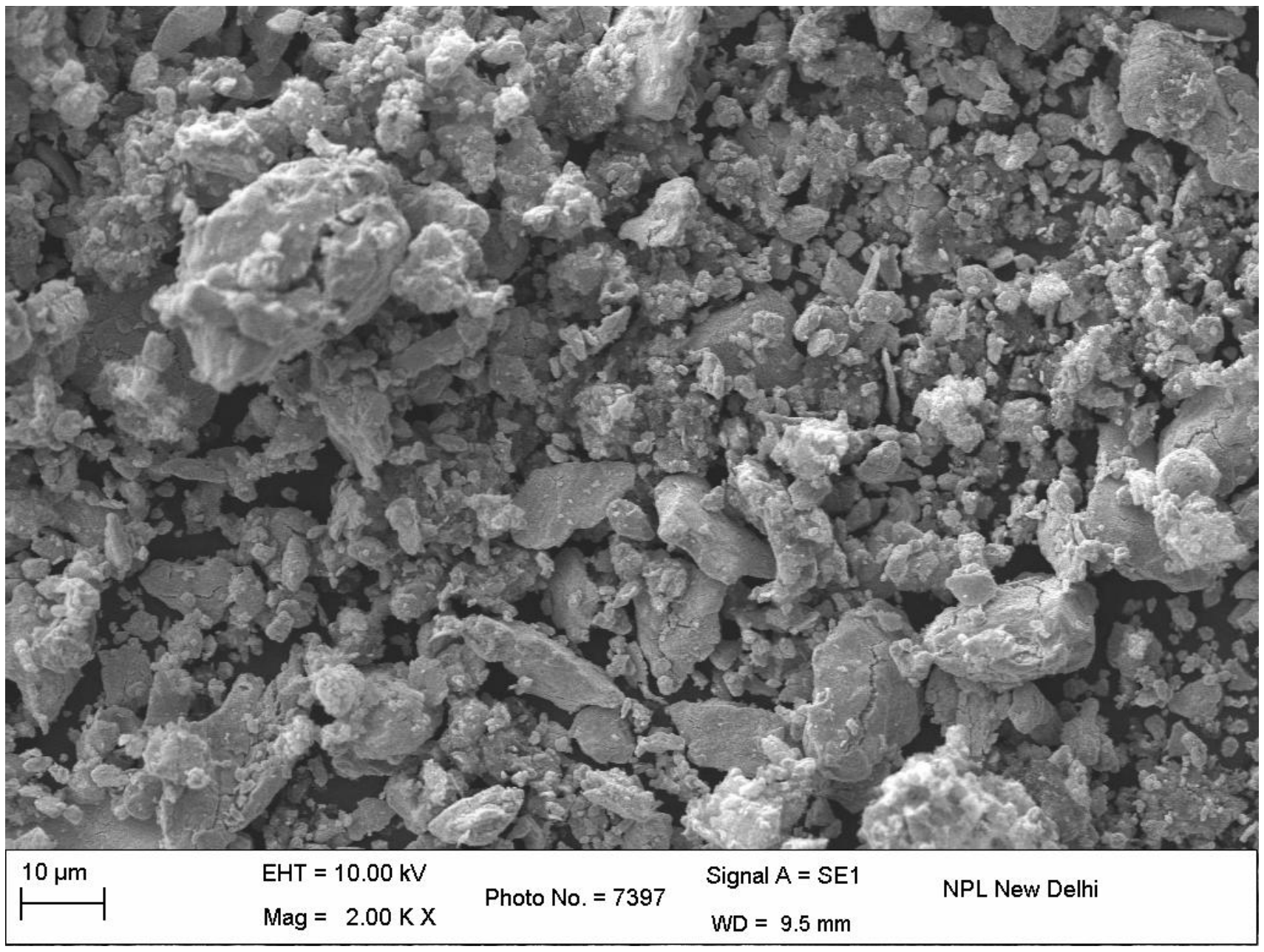

**Figure 6**

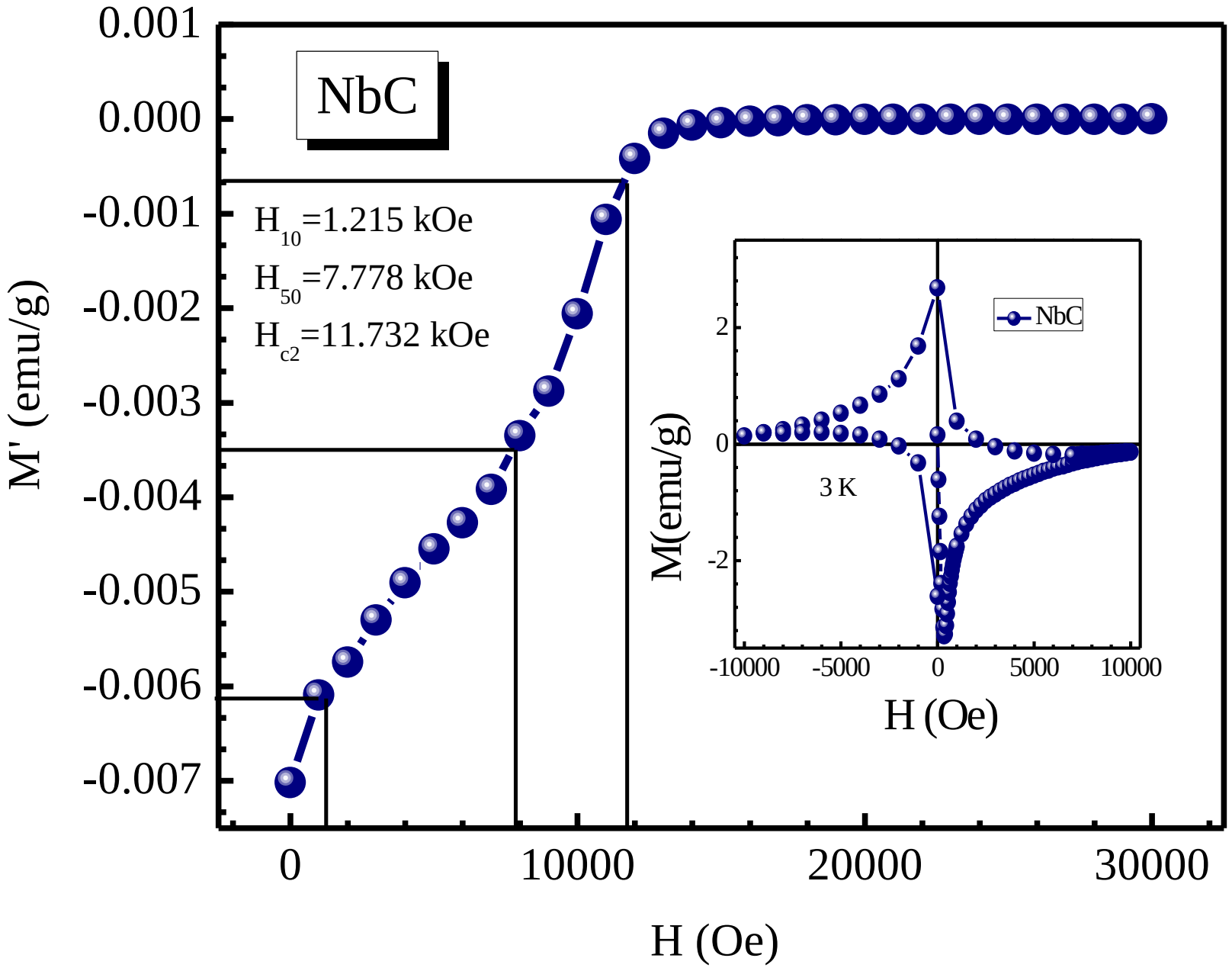